\documentclass[aps,prl,superscriptaddress,twocolumn,floatfix,a4paper]{revtex4}

\usepackage{graphicx,graphics,epsfig}   
\usepackage{dcolumn}    
\usepackage{bm}         
\usepackage{amsmath}    
\usepackage{verbatim}   
\usepackage{color}      
\usepackage{subfigure}  
\usepackage{times,natbib}
\usepackage{amsmath,amsfonts,amssymb,graphics,graphics,color,times}

\usepackage{latexsym}
\usepackage{amsmath}
\usepackage{amssymb}
\usepackage{amsfonts}
\usepackage{amsthm}
\usepackage{mathrsfs}
\usepackage{color,verbatim,graphics}
\usepackage{psfrag}
\DeclareMathAlphabet{\mathrsfs}{U}{rsfs}{m}{n}
\DeclareMathAlphabet{\mathpzc}{OT1}{pzc}{m}{it}
\DeclareMathAlphabet{\matheus}{U}{eus}{m}{n}
\DeclareMathAlphabet{\mathbbold}{U}{bbold}{m}{n}

\newcommand{\compl}{\begin{picture}(8,8)\put(0,0){C}\put(3,0.3){\line(0,1){7}}\end{picture}}

\newcommand{\ba}{\begin{eqnarray}}
\newcommand{\ea}{\end{eqnarray}}
\newcommand{\ban}{\begin{eqnarray*}}
\newcommand{\ean}{\end{eqnarray*}}
\newcommand{\Tr}{\operatorname{Tr}}

\newcommand{\ket}[1]{|#1\rangle}
\newcommand{\bra}[1]{\langle#1|}

\newcommand{\one}{\mathbbold{1}}

\begin{document}

\title{Disproving the Peres conjecture: Bell nonlocality from bound entanglement}

\author{Tam\'as V\'ertesi}
\affiliation{Institute for Nuclear Research, Hungarian Academy of Sciences,
H-4001 Debrecen, P.O. Box 51, Hungary}
\author{Nicolas Brunner}
\affiliation{D\'epartement de Physique Th\'eorique, Universit\'e de Gen\`eve, 1211 Gen\`eve, Switzerland}

\begin{abstract} 
Quantum entanglement plays a central role in many areas of physics, from quantum information science to many-body systems. In order to grasp the essence of this phenomenon, it is fundamental to understand how different manifestations of entanglement relate to each other. In 1999, Peres conjectured that Bell nonlocality is equivalent to distillability of entanglement. 
The intuition of Peres was that the non-classicality of an entangled state, as witnessed via Bell inequality violation, implies that pure entanglement can be distilled from this state, hence making it useful for most quantum information protocols. Subsequently, the Peres conjecture was shown to hold true in several specific cases, and became a central open question in quantum information theory. Here we disprove the Peres conjecture by showing that an undistillable bipartite entangled state---a bound entangled state---can nevertheless violate a Bell inequality. This shows that Bell nonlocality implies neither entanglement distillability, nor non-positivity under partial transposition, thus clarifying the relation between three fundamental aspects of entanglement. Finally, our results lead to a novel application of bound entanglement for device-independent randomness certification.
\end{abstract}

\maketitle

\section{Introduction}

The predictions of quantum theory are incompatible with any physical model that satisfies a natural principle of locality, as shown by Bell \cite{bell,review}. Specifically, the correlations obtained by performing local measurements on an entangled quantum state violate an inequality, Bell's inequality, which is satisfied by all local correlations. Understanding the link between entanglement and Bell nonlocality is a longstanding and challenging problem \cite{horo_RMP,review}. While the observation of Bell inequality violation implies the presence of entanglement, it is still not known whether all entangled states can lead to Bell inequality violation. 

While nonlocality turns out to be a generic feature of all entangled pure states \cite{gisin,PR92}, the situation is however much more complex for mixed states. There exist mixed entangled states which are local, as they admit a local hidden variable model \cite{werner}, even for the most general type of non-sequential measurements \cite{barrett}. Nevertheless, it turns out that certain local entangled states can violate a Bell inequality when a more general scenario is considered. If pre-processing by local operations and classical communication (LOCC) is performed prior to the local measurements, the 'hidden nonlocality' of some local entangled states can be revealed \cite{popescu,masanes,flavien}. Alternatively, when several copies of the state can be jointly measured in each run of the Bell test, nonlocality can be super-activated \cite{palazuelos,super}. Finally, the nonlocality of certain local entangled states can be revealed by placing several copies of the state in a quantum networks \cite{dani,sen}.

In the most general case, an arbitrary number of copies of the state can be pre-processed by LOCC operations before performing the Bell test. Hence the problem becomes intimately related to entanglement distillation \cite{peres96}. A bipartite entangled state is said to be distillable if, from an arbitrary number of copies, it is possible to extract pure entanglement by LOCC \cite{ED}. It thus follows that any entangled state that is distillable can lead to Bell inequality violation.

There exist however entangled states which are not distillable, so-called 'bound entangled' states \cite{BE}, shown to be relevant for instance in quantum many-body systems \cite{patane,toth,ferraro}. Hence the phenomenon of entanglement displays a form of irreversibility. On the one hand, entanglement is required to produce a bound entangled state, i.e. the state cannot be produced via LOCC. On the other hand, no pure entanglement can ever be extracted from a bound entangled state by LOCC. This leads naturally to the question of whether bound entangled states can also violate a Bell inequality. In 1999, Peres \cite{peres99} first discussed this problem and conjectured that bound entanglement can never lead to Bell inequality violation. 
Originally, the conjecture was formulated using the notion of partial transposition \cite{PPT}, one of the most useful tools for characterizing entanglement \cite{horo96}, directly related to symmetry under time reversal and to distillability of entanglement \cite{BE}. Specifically, Peres suggested that entangled quantum states with positive partial transpose (PPT) \cite{PPT}, and hence no distillable entanglement, can never give rise to nonlocality. An alternative formulation of the conjecture is that any entangled state that leads to Bell inequality violation must be non-positive under partial transposition (NPT). Peres' intuition was that distillability of entanglement is equivalent to nonlocality, that is, the violation of a Bell inequality by measurements on a quantum state necessarily implies that some pure entanglement can be distilled out of this state. 

In recent years, an intense research effort has been devoted to this problem. Several works provided evidence in favor of the Peres conjecture, showing that the violation of important classes of Bell inequalities implies distillability \cite{Toni,Lluis} and negativity under partial transposition \cite{WW,Dani}. On the other hand, weaker versions of the conjecture were disproven, in the multipartite case \cite{VB,Dur,augusiak}, and more recently considering the notion of quantum steering \cite{pusey,paul,tobias}. However, Peres' original conjecture remained open, and has become known as one of the main conjectures in quantum information theory. Solving this problem is thus an important challenge as it would lead to a deeper understanding of how different manifestations of the phenomenon of entanglement relate to each other (see Fig.1).

Here, we disprove the original Peres conjecture. Specifically, we present a bipartite entangled state which is PPT, hence bound entangled, but which can nevertheless violate a Bell inequality. This shows that Bell nonlocality is fundamentally different from both entanglement distillability and non-positivity under partial transposition. Finally, we show that bound entanglement can be useful in nonlocality-based quantum information protocols, in particular for device-independent randomness certification \cite{colbeck,pironio}.

\begin{figure}[t!]
\includegraphics[width = \columnwidth]{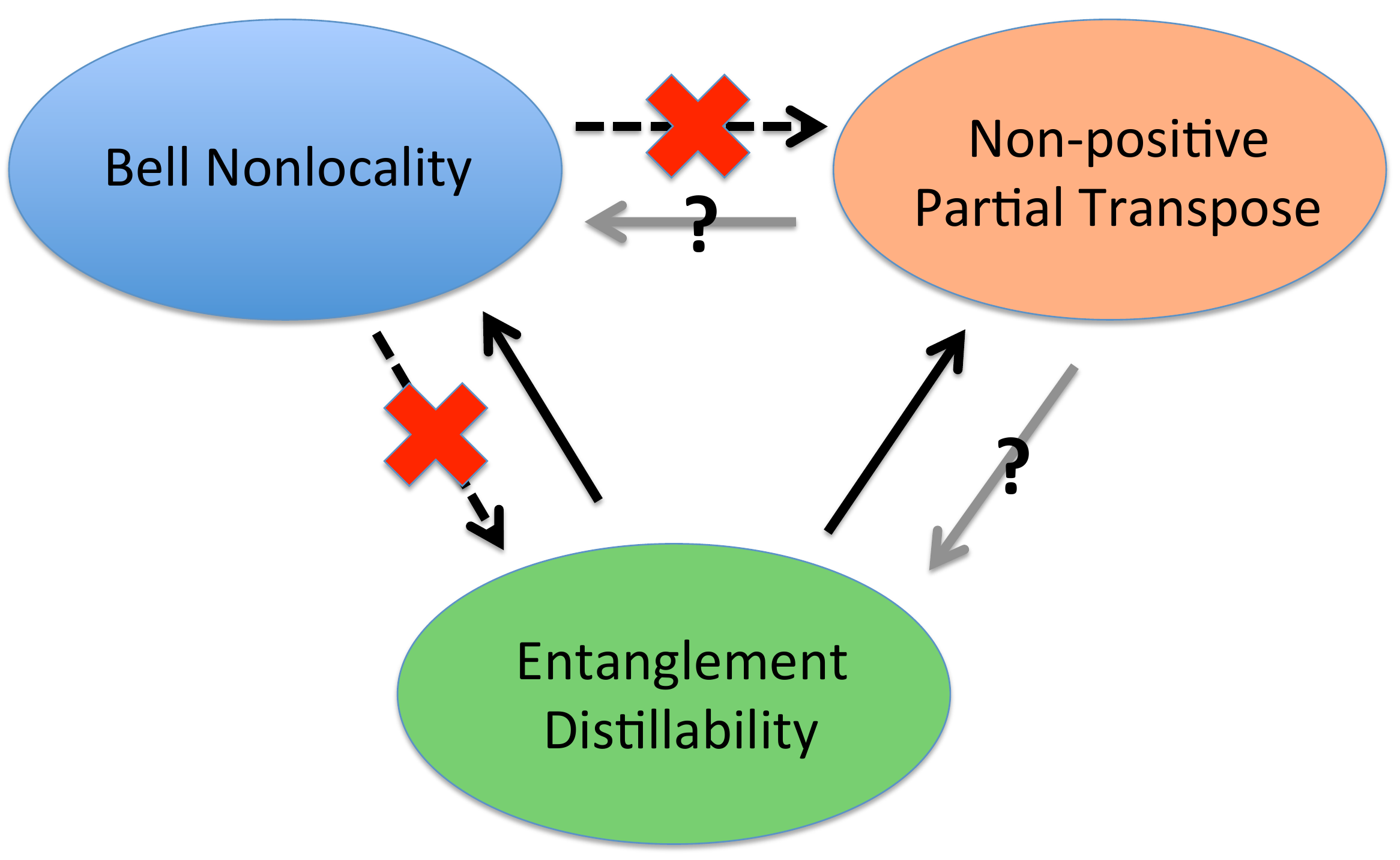}
    \caption{{\bf Relation between different fundamental manifestations of quantum entanglement.} Bell nonlocality, non-positivity under partial transposition, and entanglement distillability represent three facets of the phenomenon of entanglement. Understanding the connection between these concepts is a longstanding problem. It is well-known that entanglement distillability implies both nonlocality \cite{Lluis} and non-positive partial transpose \cite{BE}. Peres \cite{peres99} conjectured that nonlocality implies non-positivity under partial transposition and entanglement distillability; hence represented by the dashed arrows. The main result of the present work is to show that this conjecture is false, as indicated by the red crosses. To complete the diagram, it remains to be seen whether negative partial transpose implies distillability, one of the most important open questions in entanglement theory \cite{NPT_BE1,NPT_BE2}. If this conjecture turns out to be false, it would remain to be seen whether non-positive partial transpose implies Bell nonlocality.}
\label{Fig1}
\end{figure}

\section{Results}

{\bf \emph{Bound entangled state.}} We start by constructing the bound entangled state, and will later show that it violates a simple Bell inequality. We consider a state of two qutrits, i.e. of local Hilbert space dimension $d=3$. Note that there are no PPT entangled states for qubit-qubit and qubit-qutrit systems \cite{horo96}. Specifically, we consider an entangled state of the form
\begin{equation}
\varrho=\sum_{i=1}^4\lambda_i\ket{\psi_i}\bra{\psi_i}.
\end{equation}
The eigenvalues are $\lambda = \left( \frac{3257}{6884}, \frac{450}{1721}, \frac{450}{1721}, \frac{27}{6884}Ê\right)$, and the eigenvectors $\ket{\psi_i}  \in \compl^3 \otimes \compl^3$ are given by
\begin{align}
\label{psi}
\ket{\psi_1}&= \frac{1}{\sqrt 2}\left(\ket{00}+\ket{11}\right) \nonumber\\
\ket{\psi_2}&= \frac{a}{12}\left(\ket{01}+\ket{10}\right)+\frac{1}{60}\ket{02}-\frac{3}{10}\ket{21} \\
\ket{\psi_3}&= \frac{a}{12}\left(\ket{00}-\ket{11}\right)+\frac{1}{60}\ket{12}+\frac{3}{10}\ket{20}\nonumber \\
\ket{\psi_4}&= \frac{1}{\sqrt 3}\left(-\ket{01}+\ket{10}+\ket{22}\right) \nonumber,
\end{align}
where $a=\sqrt{\frac{131}{2}}$. The state $\varrho$ is part of a family of states recently discussed in \cite{tobias}. Importantly the above choice of eigenvalues and eigenvectors ensures that the state $\varrho$ is invariant under the partial transposition map \cite{PPT}, i.e. $\text{PT}(\varrho) = (\one \otimes T_B) (\varrho) = \rho$, where $T_B$ denotes the transposition operation on the second subsystem. This ensures that the state $\varrho$ is PPT, i.e. $\text{PT}(\varrho) \succeq 0$, and therefore undistillable \cite{BE}.

{\bf \emph{Bell inequality violation.}} Nevertheless, the state $\varrho$ is entangled, hence bound entangled, as it can lead to Bell inequality violation. To prove this, we consider a Bell test with two distant observers, Alice and Bob. Alice chooses between three measurement settings, $x \in \{0,1,2\}$, and Bob among two settings, $y \in \{0,1\}$. Alice's settings yield a binary outcome, $a \in \{0,1\}$. Bob's first setting ($y=0$) has a ternary outcome, $b \in \{0,1,2\}$, and his second setting ($y=1$) is binary, $b \in \{0,1\}$. The experiment is thus characterized by the joint probability distribution $p(ab|xy)$. These statistics can be reproduced by a local model if they admit a decomposition of the form:
\ba \label{local} p(ab|xy) = \int  d\lambda  \mu(\lambda) p(a | x\lambda)p(b | y\lambda)\ea
where $\lambda$ represents the shared local variable distributed according to the density $\mu(\lambda)$. For the Bell test considered here, all statistics of the above form satisfy the Bell inequality \cite{allCHSH}:
\ba\label{Inew}
 I &=& -p_A(0|2)-2p_B(0|1)-p(01|00)-p(00|10)   \nonumber \\  & & +p(00|20) +p(01|20)+p(00|01) \\ & & +p(00|11)+p(00|21) \leq 0, \nonumber
\ea
where $p_A(a|x)$ and $p_B(b|y)$ denote Alice's and Bob's marginal distributions. Hence a violation of the above inequality, i.e. $I>0$, implies the presence of nonlocality. 

In particular, this can be achieved by performing judiciously chosen local measurements on the bound entangled state $\varrho$. The local measurement operators, acting on $\compl^3$, are denoted $M_{a|x}$ for Alice and $M_{b|y}$ for Bob. 
The measurement operators of Alice are rank-1 real-valued projectors $M_{0|x}=|A_x\rangle\langle A_x|$, with
\ba \ket{A_0} &=& -p \ket{0} + \sqrt{3}p \ket{1} + \sqrt{1-4p^2} \ket{2}  \nonumber \\
\ket{A_1} &=& 2p \ket{0} + \sqrt{1-4p^2} \ket{2}   \\ \nonumber
\ket{A_2} &=& -p \ket{0} - \sqrt{3}p \ket{1} + \sqrt{1-4p^2} \ket{2}
\ea
where $p=1/5$. We have that $M_{1|x} = \one - M_{0|x}$, where $\one$ denotes the identity operator in $\compl^3$. Bob's first measurement is given by $M_{b|0}=|B_0^b\rangle\langle B_0^b|$ (for $b=0,1$) by
\ba \ket{B_0^0} &=& \sqrt{\frac{2}{3}} \ket{1} + \frac{1}{\sqrt{3}} \ket{2} \nonumber \\ 
\ket{B_0^1} &=&  -\frac{1}{\sqrt{2}} \ket{0} -\frac{1}{\sqrt{6}} \ket{1} + \frac{1}{\sqrt{3}} \ket{2}
\ea
and $M_{2|0}= \one - M_{0|0}-M_{1|0}$. For Bob's second setting, we take $M_{0|1}= \ket{2}\bra{2}$ and $M_{1|1}= \one - M_{0|1}$. The resulting statistics is given by the probability distribution 
\ba \label{stat} p(ab|xy)= \Tr(\varrho M_{a|x} \otimes M_{b|y}). \ea
These statistics do not admit a decomposition of the form \eqref{local}, as they lead to a violation of the Bell inequality \eqref{Inew}, given analytically by
\begin{align}
I_{\varrho}&=\frac{-3386 + 18\sqrt{42} - 5\sqrt{131} + 45\sqrt{5502}}{43025} \\ \nonumber 
&\simeq 2.63144 \times 10^{-4} \nonumber.
\end{align}
This proves that a bipartite bound entangled state can give rise to nonlocality, thus disproving the Peres conjecture (see Fig. 2).

To derive this result, we followed a numerical optimization method based on semi-definite programming (SDP) \cite{SDP}, briefly outlined in the Methods section. The construction described above is however analytical, and was reconstructed from the output of the optimization procedure. In fact, slightly higher Bell violations, up to $I_{PPT}=2.6526 \times 10^{-4}$, could be found numerically for two-qutrit PPT states. Moreover, using the SPD techniques of Ref. \cite{moroder}, an upper bound on the largest possible violation obtainable from PPT states is here found to be $I_{PPT}^{max} < 4.8012\times 10^{-4}$, hence leaving the possibility open for a slightly higher violation using PPT states of arbitrary Hilbert space dimension.

Finally, it is worth pointing out that our result implies that the set of PPT states violating Bell inequality \eqref{Inew} is of non-zero measure. Although the Bell violation we observe is small, it is nevertheless finite. Hence it follows that any state obtained by adding a sufficiently small (but finite) amount of an arbitrary separable state to the bound entangled state $\varrho$, will also violate the Bell inequality. As the set of separable states is of non-zero measure, the result follows.

\begin{figure}[t!]
\includegraphics[width = \columnwidth]{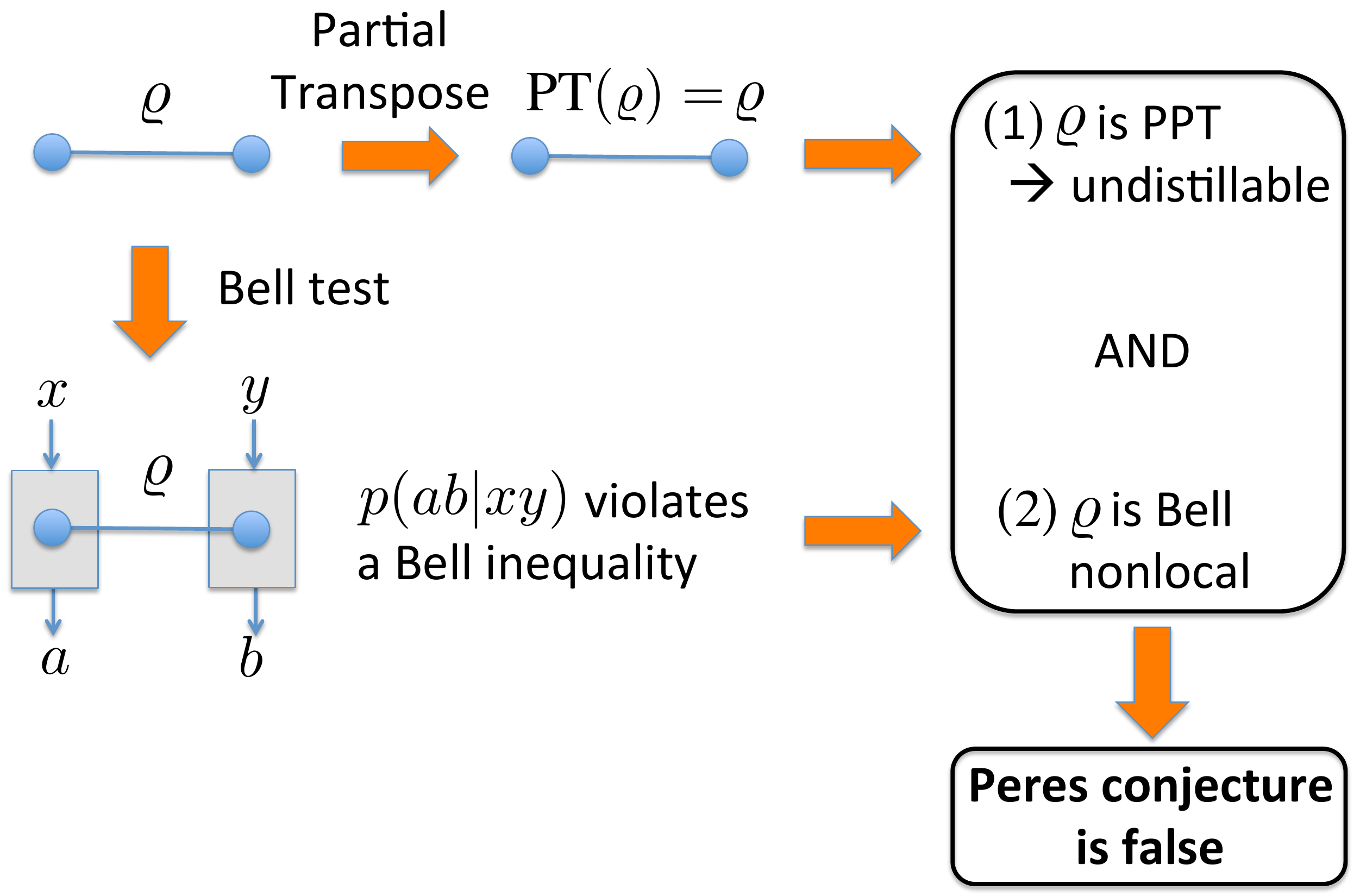}
    \caption{{\bf Building a counter-example to the Peres conjecture.} In order to disprove the conjecture, we construct a quantum state $\varrho$ with the following properties: (1) $\varrho$ is positive under partial transposition (PPT), and (2) $\varrho$ is Bell nonlocal. Property (1) follows here from the fact that $\varrho $ is invariant under partial transposition, and implies that $\varrho$ cannot be distilled. Property (2) follows from the fact that the statistics resulting from local measurements on $\varrho$ violate a simple Bell inequality.}
\label{Fig1}
\end{figure}

{\bf \emph{Randomness certification.}} The fact that a bound entangled state can violate a Bell inequality suggests potential applications in quantum information processing, in particular in nonlocality-based tasks. Here we consider randomness expansion based on Bell nonlocality \cite{pironio,colbeck}, in which true quantum randomness can be certified without relying on a detailed knowledge about the functioning of the devices used in the protocol. The security of the protocol is therefore called 'device-independent'. Following the techniques of Ref. \cite{olmo,jd}, we obtained lower bounds on the amount of randomness that can be certified from the nonlocal statistics. The randomness is captured by the probability of guessing the outcome of Bob's (or Alice's) measurement $p_g(b|y)$. Note that this guessing probability is computed by a maximization over all possible realizations that are compatible with the observed data $p(ab|xy)$, see \cite{olmo,jd} for details. To quantify the randomness it is useful to consider the min-entropy, $H_{min}=-\log_2 p_g(b|y)$, which represents the number of random bits that can be extracted per run of the Bell test (using adequate post-processing) from Bob's measurement setting $y$. The results, summarized in Table 1, show that randomness can indeed be certified using a bipartite bound entangled state. Note that in practice, implementing such a protocol would be extremely challenging, as the Bell violation is very small, and hence very sensitive to noise. 

Finally, it would be interesting to see if bound entanglement is useful for other quantum information protocols based on nonlocality. First, given its usefulness in quantum key distribution (QKD) \cite{jonathan}, it would be interesting to see if bound entanglement could also be used to establish a secret key in the context of device-independent QKD \cite{acin07}. Second, our bound entangled state could be useful in communication complexity, a task which is strongly connected to quantum nonlocality. Using the techniques of Ref. \cite{zukowski} (see also \cite{epping}), it should be possible to construct a communication complexity problem for which bound entanglement helps reducing the amount of communication compared to classical resources.

 \begin{table}[t!]
\centering
\begin{tabular}{|c|c|c|}
\hline
Bell violation $I_{PPT}$ & $H_{min}$ ($y=0$) & $H_{min}$ ($y=1$) \\
  \hline\hline
$2.6314 \times 10^{-4}$   &   $4.2320 \times 10^{-4}$  &   $3.6191 \times 10^{-4}$ \\ \hline
$2.6526 \times 10^{-4}$  &    $4.2310 \times 10^{-4} $  &  $3.6530 \times 10^{-4}$  \\ 
\hline
\end{tabular}
\caption{{\bf Device-independent randomness certification using bound entangled states.} We describe here the randomness, as quantified via the min-entropy $H_{min}$ (see main text), from the statistics of the Bell experiment. The values represent lower bounds on $H_{min}$ (obtained at the third level of the SDP hierarchy, see \cite{olmo,jd}) for the outcome of Bob's measurements ($y=0,1$). The first line corresponds to our analytical construction, while the second line is for the statistics corresponding to the largest violation we found using a PPT state. Surprisingly, it turns out that no randomness can be extracted from the outcome of Alice's measurements.}
\end{table}

\section{Discussion}

To summarize, we have shown that bipartite bound entangled states can lead to Bell inequality violation, thus disproving the long-standing conjecture of Peres. This represents significant progress in our understanding of the relation between entanglement and Bell nonlocality, demonstrating in particular that nonlocality does not imply non-positivity under partial transposition or entanglement distillability (see Fig. 1). The main open question now is whether all bound entangled state can give rise to Bell inequality violation, which would imply that entanglement and nonlocality are basically equivalent. From a more applied perspective, we also showed that bound entanglement can be useful in nonlocality-based quantum information tasks, in particular device-independent randomness certification.

\section{Methods}

\emph{Numerical method.} Consider a Bell inequality of the form
\ba I=\sum_{a,b,x,y}c_{ab|xy}p(ab|xy)\leq L \ea 
with real coefficients $c_{ab|xy}$ and local bound $L$. To find efficiently violations of such an inequality for PPT states of local Hilbert space dimension $d$, we use the following SDP procedure:
 
\begin{enumerate}
\item Generate randomly local measurement operators $M_{a|x}$ and $M_{b|y}$ acting on $\compl^d$.
    \item Construct the Bell operator: \ba\mathcal{B}=\sum_{a,b,x,y} c_{ab|xy}M_{a|x}\otimes M_{b|y}. \ea
    \item Maximize $\text{Tr}(\mathcal{B} \rho)$ subject to $\rho\succeq0$, $\text{PT}(\rho)\succeq0$, $\text{Tr}(\rho)=1$. This is an SDP, which returns an optimal state $\rho$ corresponding to the optimal Bell value $I_Q=\text{Tr}(\mathcal{B}\rho)$.
    \item Optimize the measurement operators $M_{a|x}$ for fixed $M_{b|y}$ and $\rho$ (given in step~3). This is again an SDP since $I_Q=\text{Tr}(\mathcal{B}\rho)=\sum_{a,x}\text{Tr}{(M_{a|x}F_{a|x})}$, where $F_{a|x}=\sum_{b,y}{c_{ab|xy}\text{Tr}_B(\rho \one \otimes M_{b|y})}$ are fixed $d \times d$ matrices. Hence, we maximize $I_Q=\sum_{a,x}\text{Tr}{(M_{a|x}F_{a|x})}$ subject to $\sum_a{M_{a|x}}=\one$ for all $x$ and $M_{a|x}\succeq0$ for all $(a,x)$. Thereby, we obtain Alice's optimal measurement operators $M_{a|x}$.
    \item Similarly, we obtain the optimal measurements for Bob $M_{b|y}$, for fixed $\rho$ and $M_{a|x}$.
    \item Repeat steps 3-5 until convergence of the objective value $I_Q$.
\end{enumerate}

\emph{Acknowledgements.} The authors acknowledge financial support from the J\'anos Bolyai Programme, the OTKA (PD101461), the T\'AMOP-4.2.2.C-11/1/KONV-2012-0001 project and from the Swiss National Science Foundation (grant PP00P2\_138917 and QSIT), SEFRI (COST action MP1006) and the EU DIQIP.

\end{document}